\documentclass[]{aa}
\usepackage{graphicx}
\usepackage{natbib}
\usepackage{txfonts}
\bibpunct{(}{)}{;}{a}{}{,}

\begin{document}

\title{Interstellar $^{12}$C/$^{13}$C from CH$^+$ absorption lines:
  Results from an extended survey\thanks{Based on observations
  obtained with Feros at the ESO/MPI 2.2m telescope, La Silla, Chile
  (proposal No. 076.C-0431(A)) and on observations obtained with UVES
  at the ESO Very Large Telescope, Paranal, Chile (proposal
  No. 076.C-0431(B)).}  }

\titlerunning{Interstellar $^{12}$C/$^{13}$C}

\author{O. Stahl
  \inst{1}
  \and  S. Casassus \inst{2}  \and T. Wilson\inst{3,4}}

\institute{ZAH, Landessternwarte K\"onigstuhl, 69117 Heidelberg, Germany.
  \email{O.Stahl@lsw.uni-heidelberg.de}
  \and
  Departamento de Astronom\'ia, Universidad de Chile,
  Casilla 36-D, Santiago, Chile.
  \email{simon@das.uchile.cl}
  \and
  ESO, Karl-Schwarzschild-Str. 2, 85748 Garching bei M\"unchen, Germany.
    \email{twilson@eso.org}
  \and
  Max-Planck-Institut f\"ur Radioastronomie, Postfach 2024, D-53010 Bonn, Germany}

\offprints{O. Stahl} 

\date{Received / Accepted} \abstract{ The $^{12}$C/$^{13}$C isotope
  ratio in the interstellar medium (ISM), and its evolution with time,
  is an important tracer of stellar yields. Spatial variations of this
  ratio can be used to study mixing in the ISM\@.}  {We want to
  determine this ratio and its spatial variations in the local ISM
  from CH$^+$ absorption lines in the optical towards early-type
  stars. The aim is to determine the average value for the local ISM
  and study possible spatial variations. } {We observed a large number
  of early-type stars with Feros to extend the sample of suitable
  target stars for CH$^+$ isotope studies.  The best suited targets
  were observed with Uves with higher signal-to-noise ratio and spectral
  resolution to determine the isotope ratio from the interstellar
  CH$^+$ lines. This study significantly expands the number of
  $^{13}$CH$^+$ detections. }  {We find an average ratio of $\langle R
  \rangle = 76.27 \pm 1.94$ or, for $f = 1/R$, $\langle f \rangle =
  (120.46 \pm 3.02) \times 10^{-4}$. The scatter in $f$ is
  $6.3\times\sigma(\langle f \rangle)$. This findings strengthens the
  case for chemical inhomogeneity in the local ISM, with important
  implications for the mixing in the ISM\@. Given the large scatter,
  the present-day value in the ISM is not significantly larger than
  the solar value, which corresponds to the local value 4.5\,Gyr ago.}
     {} 
  
\maketitle

\section{Introduction}

The $^{12}$C/$^{13}$C ratio is a cornerstone of models of the nuclear
history of our galactic interstellar matter (ISM), that is, the
Galactic Chemical Evolution.  Low and intermediate mass stars increase
(during the AGB phase) the $^{12}$C/$^{13}$C ratio to $\sim300$, while
stars more massive stars (undergoing CNO processing) increase the
ratio to $^{12}$C/$^{13}$C$\sim$3. A measurement of the
$^{12}$C/$^{13}$C ratio in the ISM thus provides important information
about the relative importance of stars of different mass to the
chemical enrichment of the interstellar matter.

The $^{12}$C/$^{13}$C ratio in the solar system is 89
\citep{1989GeCoA..53..197A}. Models of chemical evolution predict a
decrease in the $^{12}$C/$^{13}$C ratio with time for a given
galacto-centric distance, and a decrease with galacto-centric distance
in the Galaxy \citep{1996A&A...309..760P}.  Such models assume that
mixing in the ISM is complete and restricted to material at a given
galacto-centric distance (azimuthal mixing). A comparison of results
between our Galaxy and other galaxies will give important data for the
nuclear processing history \citep[see,
e.g.,][]{2000emw..conf..505T,2001coev.conf..233P}

The $^{12}$C/$^{13}$C ratio has been determined mostly from
radio-astronomical measurements of CO and H$_2$CO \citep[see
e.g.,][]{1994ARA&A..32..191W} and millimeter observations
of CN \citep{2002ApJ...578..211S,2005ApJ...634.1126M}.

These results may be affected by interstellar chemistry in different
ways, however: Chemical fractionation (which results from different
zero-point energies of the isotopomers) enriches molecules in $^{13}$C
\citep{1976ApJ...205L.165W} and thus lowers the ratio, selective
dissociation destroys the rarer species more and thus raises the
ratio. The latter effect is due to self-shielding of the more abundant
isotopemer, which therefore survives photo-destruction better than the
less abundant counterpart.

The molecule CH$^{+}$ is thought to be produced only in hotter parts
of photo-dissociation regions and is a rare species, so is not
affected by either chemical fractionation or selective
dissociation. In addition, the optical CH$^+$ lines have typically
only small saturation effects. This molecule, which can be observed in
optical absorption lines from the ground, is therefore considered to
provide the most secure $^{12}$C/$^{13}$C ratios \citep[see
  e.g.][]{1976ApJ...205L.165W}. We caution, however, that the
formation process of CH$^+$ is probably not completely understood
\citep{1993A&A...269..477G}.

Two systems of CH$^{+}$ lines are strong enough to be easily observed
from the ground. In addition, the absorption lines from
$^{12}$CH$^{+}$ at 4232 and 3957 \AA\ are sufficiently shifted from
their $^{13}$CH$^{+}$ component to allow unblended measurements,
provided the velocity profiles in the intervening cloud are narrow
enough. If both are measured simultaneously, we can make use of the
fact that the $^{13}$CH$^{+}$ line is blue-shifted by 0.26 \AA\
relative to the $^{12}$CH$^{+}$ at 4232 \AA\ but red-shifted by 0.44
\AA\ relative to the $^{12}$CH$^{+}$ line at 3957 \AA\@. This way, a
chance superposition with another cloud along the line-of-sight can be
excluded. In addition, the two simultaneous measurements help to eliminate 
systematic effects. 

This property has been used e.g.\@ by \citet{1992A&A...254..327S} to check
that there is no accidental overlap of a weak velocity component of
$^{12}$CH$^{+}$ at $-$18.8 km sec$^{-1}$ from the deepest
$^{12}$CH$^{+}$ absorption line with the $^{13}$CH$^{+}$ line. Nearly
all previous data were taken at 4232 \AA\ only.

These measurements in the optical require, however, observations with
high spectral resolution and signal/noise ratio. Most of the published
CH$^+$-measurements have therefore been obtained with very bright
background stars at relatively small distances from the Sun.

The availability of efficient high-resolutions spectrograph at
8m-class telescopes in recent years has improved the situation
considerably \citep[][in the following 
paper\,I]{2005A&A...441..181C}. The observations are hampered,
however, by the scarcity of suitable published candidate stars with
sufficiently strong and narrow CH$^+$ absorption lines.  In this paper
we describe new observations which have been obtained at targets
specifically selected for this purpose.

The outline of the paper is as follows: Section~\ref{sec:obs}
describes our observations. Section~\ref{sec:model} gives details on
our procedure to measure the isotopic ratio $R$. The results are
discussed in section~\ref{sec:results}, both the average properties of
the sample and notes about individual targets. Section~\ref{sec:conc}
discusses the results and concludes the paper.

\section{Observations and reduction} \label{sec:obs}

\subsection{Pre-selection with Feros}

Many of the published CH$^+$-measurements are hampered either by too broad or
complex line profiles of the interstellar lines. The isotope shift
is very small, so that for most sight-lines (with relatively broad
absorption lines) the isotopic lines partly overlap, which introduces
considerable errors to the isotope ratio. Interstellar lines with
several components are even more problematic. The best lines of sight
have sharp, single absorption lines.

The stars with published CH$^+$-profiles that are suitable for
the determination of the $^{12}$C/$^{13}$C isotope ratio are exhausted
by previous studies. We therefore undertook a dedicated search for additional
targets, which might be suitable for studies of the isotope ratio.  

For this purpose we used the {\sc Feros} echelle spectrograph at the
ESO/MPI 2.2m telescope at ESO, La Silla. {\sc Feros} has a
sufficiently high spectral resolution (48\,000) to examine the CH$^+$
line profiles. In three nights in January 2006 we observed about 80
early-type stars with a B magnitude brighter than 8.0 selected from
various lists \citep{1978ApJS...38..309H,2004ApJS..151..103M}.  

The exposure times for the {\sc Feros} spectra were relatively short
(typically a few minutes), just long enough to give a reasonable
S/N-ratio near the $\lambda$4232, 3957 lines of CH$^+$. The
$^{13}$CH$^+$ component is not detected in these spectra. The {\sc Feros}
data were pipeline-reduced directly at the telescope.  From these
spectra we selected the targets which were best suited for follow-up
observations with {\sc Uves}. The main selection criteria were:

\begin{itemize}
\item sharp and symmetric CH$^+$-lines
\item high equivalent width and 
\item brightness of the target in the B band
\end{itemize}

The sharpest and deepest absorptions are
needed to separate $^{13}$CH$^+\lambda4232$ from $^{12}$CH$^+\lambda4232$. We first selected by
inspection the narrowest and single-component absorptions, with line
profiles close to a single Gaussian. We then ordered individual lines
of sight according to two quantitative indicators of sharpness. The
first indicator is the signal-to-noise ratio on the absorption line,
$\gamma = D/\sigma = W_\lambda \sqrt{F_\mathrm{c}} / \Delta = W_\lambda
10^{-0.2\mathrm{B}} / \Delta$, where $D$ is the depth of the line,
$W_\lambda = (\sqrt{\pi}/(2\sqrt{2\ln 2}) ) D \Delta / F_\mathrm{c}$ its
equivalent width, and $\Delta$ its full width at half maximum, $\sigma
\propto \sqrt{F_\mathrm{c}} $ is the noise, and $B$ is the B-band magnitude of
the background star. The second indicator, introduced to weight the
final ranking towards the sharpest lines, was simply $\gamma_\Delta =
\gamma / \Delta$. The final list of stars selected and later observed
with {\sc Uves} is given in Table~\ref{table:obs}.

Since the {\sc Feros} spectra cover a large spectral range (3600 --
9200 \AA) we also obtained information about many other interstellar
lines. A paper describing the full {\sc Feros} data set is in
preparation.

\subsection{{High S/N observations with \sc Uves}}

Observations of the selected stars with high S/N were obtained with
the {\sc Uves} echelle spectrograph at the VLT unit telescope Kueyen
at Cerro Paranal, Chile, in three nights between January 8/9 and
January 10/11, 2006. {\sc Uves} allows measuring both line systems in
the same detector setting, thus providing a means to correct for line
blending if it is apparent in the spectra.

The observations are difficult because of the need for both very high
S/N and very high spectral resolution to acquire the profile of faint
$^{13}$CH$^{+}$, against a very bright stellar continuum and close to
the much stronger $^{12}$CH$^+$ component. We need S/N ratios of at
least $\sim$10 on $^{13}$CH$^{+}$ for accurate profile fitting.  

The project requires the highest possible spectral resolution.  We
therefore used an image slicer to minimize flux losses. In addition,
the image slicer distributes the light along the slit, which improves
flat-fielding and allows longer integration times before saturation
occurs. {\sc Uves} slicer \#2 was used, which reformats an entrance opening
of 1\farcs8 $\times$ 2\farcs2 to a slit of 0\farcs44 width and a
length of 7\farcs9, which is imaged on the spectrograph entrance slit
of 0\farcs45 width. The spectral resolution in this configuration is
$\lambda$/$\Delta\lambda$ = 75\,000. The central wavelength was set to
4\,370 \AA, which gives a spectral coverage from 3\,730 to 5\,000
\AA\@.  During our observations the seeing was sometimes too good to
fill the entrance aperture of the images slicer. Therefore, for some
spectra, only one or two slices contained most of the signal, which
unfortunately decreases the gain of the image slicer.

Accurate flat-fielding is also important. Therefore a large number of
flat-fields (150) was obtained during day-time distributed along the
observing run. 

In addition, two rapidly rotating unreddened early-type stars
(HD\,10144 = $\alpha$Eri, spectral type B3Ve and HD\,108248 =
$\alpha^1$Cru, spectral type B0.5IV) were observed in order to check
for possible faint telluric features. These stars have been observed
with exactly the same instrumental settings as the target stars and
with similar S/N-ratio. We confirm there are no detectable telluric
features under the CH$^+$ absorption.

The brighter targets where exposed until about 50\% of the maximum
level allowed by the CCD detector was reached. This to typically a few
minutes for our targets.  Series of up to 50 exposures per
target were obtained to build up the required S/N-ratio. Targets which
turned out to show too complex line profiles were dropped from the
target list.  The observations are summarized in
Table~\ref{table:obs}, which also lists the mean airmass of the target
objects in the different nights. At the higher airmass values, the
range in airmass differed from the mean by about $\pm$0.1 during the
observations, and less at smaller airmass.

We used the {\sc Uves}-pipeline software for the reduction of the
spectra. In order to maximize the S/N-ratio of the extracted spectra,
all flat-fields obtained during the run were averaged. A mean
ThAr-spectrum obtained during day time was used for the calibration of
all spectra obtained in one night. After background subtraction and
flat-fielding, the spectra were extracted as 2D-spectra using a long
extraction slit covering all slices. All spectral orders of these
spectra were then merged in one 2D-spectrum, keeping the information
along the slit.  Finally, the spectra were collapsed to a 1D-spectrum
by averaging along the slit and thereby averaging the flux from all
slices.  The region around the CH$^+\lambda$3957\,\AA-line was influenced by
a detector blemish. This defect only covered two CCD lines along the
slit. In order to remove this defect, we also extracted spectra
excluding the two affected CCD lines.  The observations of of each
target were then averaged to a mean spectrum. Wavelengths are reported
in air and referred to the solar barycenter.

\begin{table}[h]
\caption{Summary of the observations.  The table columns contain the 
HD\,number, spectral type, B magnitude, average airmass, total
integrated exposure time and the number of exposures for each target.}
\begin{tabular}{lllrrr}
Target  & Spec. &  B  & Airmass & Exp. &  Number  \\
  &  &  &  &  [sec] & \\
\hline
HD10144 &  B3Ve   & 0.3  &   1.36 &    63 &    90 \\ 
HD23480 &  B6IVe  & 4.11 &   1.52 &  1100 &    55 \\ 
HD35149 &  B1V    & 4.85 &   1.16 &  3920 &    98 \\ 
HD37903 &  B1.5V  & 7.91 &   1.13 &  5760 &    24 \\ 
HD52266 &  O9V    & 7.22 &   1.15 &   606 &     3 \\ 
HD52382 &  B1Ib   & 6.64 &   1.05 &  4800 &    40 \\ 
HD53974 &  B0.5IV & 5.44 &   1.03 &  3020 &    60 \\ 
HD58510 &  B1Iab  & 6.91 &   1.04 &  5000 &    25 \\ 
HD73882 &  O9III  & 7.58 &   1.05 &  6257 &    27 \\ 
HD75149 &  B3Ia   & 5.68 &   1.09 &  5400 &    90 \\ 
HD76341 &  B1/B2Ib & 7.39 &   1.06 &  6000 &    25 \\ 
HD91452 &  B0III  & 7.66 &   1.30 &  5760 &    24 \\ 
HD92964 &  B2.5Iae & 5.59 &  1.22 &  3900 &    65 \\ 
HD108248 &  B0.5IV & 1.32 &  1.31 &    20 &    20 \\ 
\end{tabular}
\label{table:obs} 
\end{table}

\section{Model line profiles} \label{sec:model}

We model the CH$^{+}$ opacity in velocity space, as in paper\,I. But
in this work we force the same opacity profile $\tau(v)$ on both
overtones. In paper\,I we preferred to fit the two overtones
separately, and thus obtain independent measurements with which to
assess the role of systematics. Here we incorporate the prior
knowledge that both CH$^{+}\lambda\lambda$3957,4232 overtones share a
common opacity profiles, which is true because they both stem from the
same de-excited energy level \citep[the ground state, see
][Sec.1]{2005A&A...441..181C}

Requiring a common velocity profile highlights the limits of our
knowledge of the underlying continua on the wavelength scales of the
CH$^+$ line widths. The continua can be affected by detector glitches,
or atmospheric features. In the absence of instrumental artifacts or
stellar features, the two profiles should be the same. Any difference
in the fit residuals should highlight problems with the continuum
definition. 

In this paper we investigate how a simultaneous fit to both overtones
helps constrain the CH$^+$ opacity profiles, decomposed in a set of
Gaussians in velocity. This approach was attempted in paper\,I to
resolve cases when the $^{13}$CH$^+\lambda$4232 transition was
blended with the main isotope. But in paper\,I the quality of the
simultaneous fits was not satisfactory. We now explain this problem in
paper\,I as being due to our overlooking the possibility of small
shifts in the observed relative wavelengths of the two overtones,
$\Delta\lambda_\circ$. Allowing for $\Delta\lambda_\circ$ produces
satisfactory simultaneous fits to both overtones. Such wavelengths
shifts are of the order of the uncertainties in the absolute
wavelength calibration of each echelle of a few m\AA\@. From the fits to
the 11 lines of sight acquired in 2006, we find that the maximum range
of values for $\Delta\lambda_\circ$ is $3.2~10^{-3}$\,\AA, and its
$1\sigma$ scatter is $9.5~10^{-4}$\,\AA, while the resolution element is
5.3~10$^{-2}$\,\AA\@. 

The isotope ratio information in cases where the rare isotope is
blended at $\lambda$4232 cannot be recovered without the $\lambda$3957
information. The uncertainties in the baseline definition make it
impossible, using a single transition, to distinguish the rare isotope
from a detector glitch. But $\lambda$4232 blends are still useful as a
consistency checks. We incorporate the uncertainties in the baseline
definition under $\lambda$4232 through the inclusion of extra
free-parameters. The possibility of detector glitches or stellar or
atmospheric features is modeled with a correction to the baseline
under the interstellar absorption. We explored the possibilities of
two broad types of low-level baseline corrections: 1- additive
corrections, as in low-level atmospheric, stellar or CH$^+$ features,
and 2- multiplicative corrections, as in flat fielding features. The
low-level baselines were modeled using either an expansion in Legendre
polynomials or as segmented lines.  But the Legendre expansion
required high orders, and thus many free-parameters, because of the
condition of fixed endpoints imposed by continuity with the baseline
neighboring the interstellar absorption.  We obtained best results
with an additive baseline correction, composed of a segmented line
defined by the amplitude of spikes distributed uniformly in
wavelengths.

We caution that the present technique used to measure the CH$^+$
isotopic ratio suffers from the same uncertainties inherent to any
observation of absorption lines. The inferred equivalent widths depend
on the exact placement of the continuum level. The uncertain
definition of the baseline is a source of systematic
uncertainty. However we take a conservative value for the thermal
noise, with the hope of accounting for at least small variations in
the baseline definition. Our noise estimate corresponds to the
standard deviation of the residuals in the whole 1.5\,\AA\ windows
around CH$^+\lambda\lambda3957,4232$, after baseline and profile
fitting, including detector glitches and any telluric
features. Provided the baselines are realistic enough, our procedure
should indeed allow for small variations among plausible baselines.

The model we use to define the underlying continua and to parameterize
$\tau(v)$ is otherwise the same as in paper\,I. For clarity we now
itemize our new procedure: 

\begin{enumerate}

\item extract a 1.5\,\AA\ spectrum $F(\lambda)$ centered on the CH$^{+}$
  line, and fit a baseline with a $4-7^\mathrm{th}$ order Legendre
  polynomial; include weights to improve the quality of the baselines
  near the features of interest; store the RMS dispersion of the
  residuals as the spectrum's noise, $\sigma_\mathrm{F}$;  \label{it:local}

\item define a model spectrum with
  \begin{equation}
    F_\mathrm{m}(\lambda) = F_\mathrm{c_1}(\lambda) \exp(-\tau(\lambda)),
  \end{equation}
  where the line absorption opacity $\tau$ is a superposition of
  $n_\mathrm{g}=1$ to $4$ Gaussians common to each isotope, with
  $^{13}$CH$^{+}$ components sharing the parameters of the
  $^{12}$CH$^{+}$ components, except their centroids are translated by
  a velocity shift $\Delta v_\mathrm{iso}$, and their amplitude are
  scaled by a factor $f=^{13}$CH$^{+}/^{12}$CH$^{+}$, kept as a free
  parameter:

  \begin{eqnarray}
  \tau(\lambda) & =  & \sum_{i=1}^{n_\mathrm{g}} \tau^{\circ}_\mathrm{i} \exp\left( 
  -0.5 (\lambda - \lambda^{\circ}_\mathrm{i})^2 / {\sigma^{\circ}}_\mathrm{i}^2 \right) + \nonumber \\ 
& & \sum_{i=1}^{n_\mathrm{g}}  f\, \tau^{\circ}_\mathrm{i} \exp\left( -0.5 (\lambda - 
  \lambda^{\circ}_\mathrm{i} (1+\Delta
  v_\mathrm{iso}/c))^2 /  {\sigma^{\circ}}_\mathrm{i}^2 \right) ;
  \end{eqnarray} \label{it:model}

\item The wavelength shift between each isotope is taken as a free
  parameter, allowed to vary over a resolution element.

\item In cases of blends or suspected glitches we include an additive
  correction to the baseline (details are given in the description of
  results from individual lines of sights). The corrections applied to
  the continuum level derived from Step~\ref{it:local}, are parameterized
  using a segmented line defined by $n_\mathrm{spikes}$ points
  distributed in equal wavelength intervals under the whole CH$^+$
  profile. 

\item Since blends always occur under $\lambda$4232, we artificially
  increase the noise under the fundamental tone by a factor of 10,
  thereby allowing the otherwise noisier overtone to dominate the
  optimization. 

\item Optimize the model parameters by minimizing $\chi^2 = \sum_{j}
  (F(\lambda_\mathrm{j})-F_\mathrm{m}(\lambda_\mathrm{j}))^2/\sigma^2_\mathrm{F}$,
  using the noise from Step~\ref{it:local}. The optimization was
  carried with the PDL::Minuit package (based on the Minuit
  optimization library from CERN)

\item Update the noise of the stellar spectrum by replacing
  $\sigma_\mathrm{F}$ with the root mean square (RMS) dispersion of
  the residuals (the difference between the observed and model spectra
  over the whole 1.5\,\AA\ window).

\item Estimate the uncertainty in model parameters by two methods:
  \begin{enumerate}
  \item search parameter space for the $\Delta \chi^2 = 1$ contour
    defining the region enclosing 68.3\% confidence level. 
  \item estimate $1\sigma$ uncertainties from the curvature matrix of $\chi^2$
    (i.e.\ approximate to normal errors). 
  \end{enumerate} 
\end{enumerate}

\section{Results of the fits}\label{sec:results}
\subsection{Average properties of the dataset}

In the following we discuss the results of our fits for all
sight-lines.  In order to increase the number of sight-lines, we
include, in addition to the targets observed in Jan.~2006 also the
best targets from paper\,I: $\zeta$\,Oph, HD\,110432 and HD\,152235.
These stars have been observed with the same instrument setting but
were analyzed differently.  In order not to introduce any bias in our
results, we applied the fitting method described above also for these
stars. The results for $R$ and $f$ differ slightly from the results in
paper\,I, but are consistent within the errors.

\begin{figure}
\begin{center}
\includegraphics{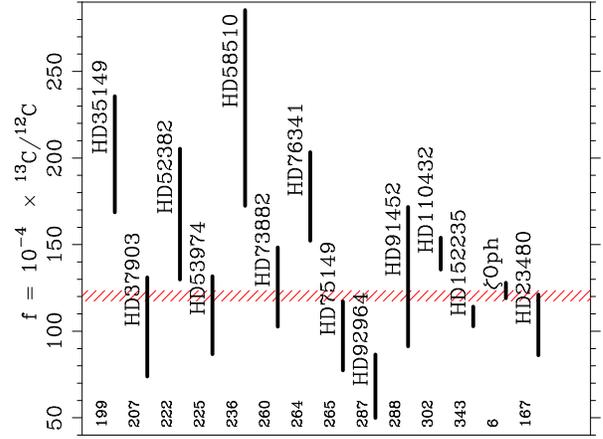}
\end{center}
\caption{A plot of individual $f$ values for the lines of sight
  studied in this work. The galactic longitude of the targets is
    also indicated. The thick lines are $\pm 1\sigma$ uncertainties
  ($2~\sigma$ in total length). The hatched box is the weighted
  average of the measurements in the figure, and should be
  representative of the local ISM. The height of the hatched rectangle
  is $\pm 1~\sigma$.  This value is $\langle f \rangle = (120.46 \pm
  3.02) \times 10^{-4}$. The corresponding average for $R$ values is
  $\langle R \rangle = 76.27 \pm 1.94$.}
\label{fig:all}
\end{figure}

The results for the fits of all targets are given in
Table~\ref{table:fits} and are plotted in Fig.~\ref{fig:all}.

\begin{table*}
\caption{Summary of the results of the fits. $f$ is given in units of
10$^{-4}$. $n_\mathrm{g}$ is the number of Gaussian components and
W$_\lambda$(4232 \AA) and W$_\lambda$(3957 \AA) are the equivalent
widths of the respective $^{12}$CH$^+$ lines, obtained by
integrating the fit solutions. Numbers in parentheses are
uncertainties in the last digit. The corresponding values for the
$^{13}$CH$^+$ components can be obtained by multiplying the values for
the $^{12}$CH$^+$ component with $f$.}
\begin{tabular}{lr@{}c@{}l@{$\pm$}r@{}c@{}lr@{}c@{}l@{$\pm$}r@{}c@{}rrrr}
Target  & \multicolumn{6}{c}{$R\pm\sigma(R)$} &   \multicolumn{6}{c}{$f\pm\sigma(f)$}  & 
$ n_\mathrm{g} $ & W$_\lambda$(4232 \AA)      & W$_\lambda$(3957 \AA) \\ 
  & \multicolumn{6}{c}{} &  \multicolumn{6}{c}{} & & [m\AA]     & [m\AA] \\ 
\hline
HD23480   & 96&.&5 & 16&.&3& 103&.&6 &  17&.&5   &1 & 14.94 (2) &  8.82 (3)    \\
HD35149   & 49&.&4 &  8&.&2  & 202&.&2 &  33&.&5  & 2 &  9.48 (3) &  5.29 (3)    \\
HD37903  &     97&.&5  & 27&.&2  &  102&.&5 &  28&.&6   & 2 & 10.05 (3) &  5.80 (4)    \\
HD52382  &     59&.&7  & 13&.&5  &  167&.&6 &  37&.&8   & 2 & 19.20 (3) & 11.01 (4)    \\
HD53974  &     91&.&5 & 18&.&8  &  109&.&2 &  22&.&5   & 2 & 11.76 (2) &  6.68 (3)    \\
HD58510  &     43&.&7 & 10&.&8  &  228&.&9 &  56&.&5    & 3 & 10.30 (3) &  5.75 (4)    \\
HD73882  &     79&.&7 & 14&.&5  &  125&.&5 &  22&.&9    & 2 & 17.45 (3) &  9.86 (4)    \\
HD75149  &    102&.&7 &21&.&0  &  97&.&3 &  19&.&9    & 2 & 10.22 (2) &  5.73 (3)    \\
HD76341  &     56&.&2 &  8&.&1  &  177&.&8 &  25&.&6    & 2 & 38.43 (3) & 23.10 (4)    \\
HD91452  &     76&.&0 &  23&.&3  & 131&.&5 &  40&.&2    & 2 &  8.75 (3) &  4.86 (4)    \\
HD92964  &    146&.&7 &  39&.&3  & 68&.&2 &  18&.&3    & 2 &  8.06 (2) &  4.46 (2)    \\
$\zeta$\,Oph  &  80&.&9 &  3&.&0  & 123&.&5 &  4&.&5    & 2 & 23.42 (1) & 13.82 (1)    \\
HD110432  &     69&.&1  &   4&.&4  & 144&.&8 &  9&.&2    & 2 & 13.74 (2) &  7.88 (2)    \\
HD152235  &     92&.&2 & 4&.&9  &  108&.&5 &  5&.&7    & 3 & 41.78 (3) & 24.88 (2)    \\
\end{tabular}
\label{table:fits} 
\end{table*}

 In paper\,I we summarized our measurements in terms of $R =
^{12}$C/$^{13}$C. But it is preferable to use $f = 1/R$ since for $f$
the noisiest factor occurs in the numerator.  We combine all $f$
measurements in a weighted average $\langle f \rangle$ by weighting
each line of sight by $1/\sigma^2$, where $\sigma$ is the
root-mean-square uncertainty on $f$. We find $\langle f \rangle =
(120.46 \pm 3.02) \times 10^{-4}$ . The weighted scatter of $f$ values
is $6.3\times\sigma(\langle f \rangle)$, which means that the scatter
is truly a property of the distribution, and not merely a result of statistical
fluctuations about  a single mean value. Indeed, applying the $\chi^2$ test for
the null hypothesis that all measurements derive from a single value,
we get that the reduced $\chi^2$ is 3.02 for 14 data points and 1 free
parameter (i.e. $\langle f \rangle$). Such a $\chi^2$ value allows us
to state that the observed distribution of $f$ values is a real
scatter with essentially 100\% confidence.

The corresponding average for $R$ values is $\langle R \rangle = 76.27
\pm 1.94$, and can be compared to the value of $R = 78.27 \pm 1.84$
published in \citet{2005A&A...441..181C}, calculated from a smaller sample.

Contrary to $f$, the isotope ratio $R = 1/f$ is not normally
distributed, so that the uncertainties on $R$ are not directly related
to confidence levels. Therefore the ISM statistics over the sample of
sight-lines must be taken on $f$. Unfortunately in paper\,I we extracted
statistics on $R$ values, rather than $f$ values. Here we have
corrected this mistake. 
The conclusions from paper I still hold, however. The positive
skewness in $R$ becomes negligible for smaller uncertainties and
larger $R$. In the paper I data the ISM scatter in $R$ turns out to be
7\% smaller than that in $f$ relative to the uncertainties on their
weighted average. The variations in $f$ towards HD~110432 and
HD~152235 are 3.4$\sigma$ (rather than 4.3$\sigma$ in $R$), and are
3.5$\sigma$ towards HD~110432 and HD~170740 (rather than
3.2$\sigma$).

\subsection{Notes on individual objects}

\subsubsection{\object{HD\,23480}}

\begin{figure*}
  \centering
  \includegraphics{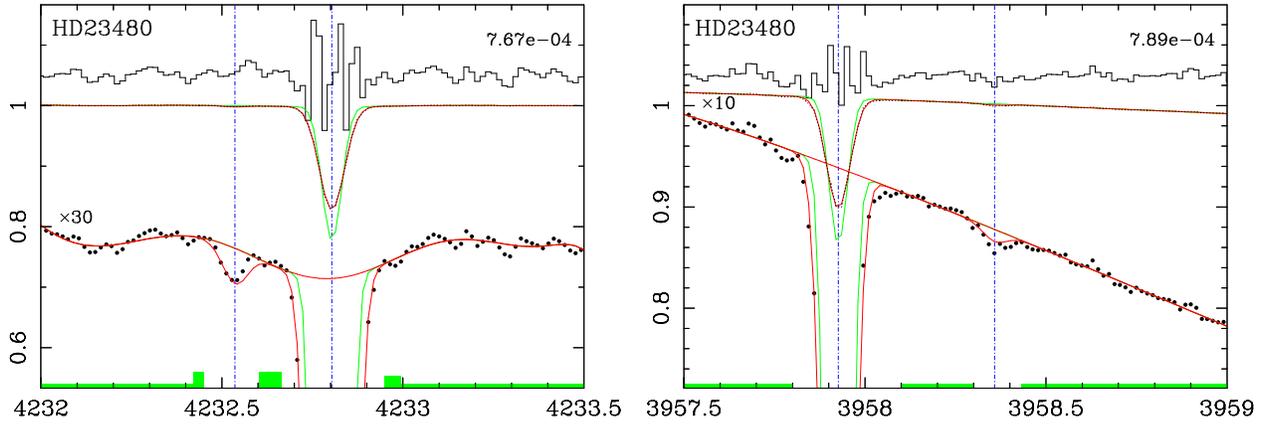}
  \caption{Spectra of HD\,23480. Wavelengths are in \AA, and flux
densities in arbitrary units. The individual Gaussian components
comprising the fit, prior to folding with the instrumental response,
are shown only on $^{12}$CH$^+$ as light gray lines, but are omitted
from $^{13}$CH$^+$ for clarity. The vertical dashed lines mark the
line centroids, at the average of the Gaussian centroids weighted by
their equivalent widths. The units of the $y$-axis are arbitrary, and
are scaled so that the median of the object spectrum is unity. Also
shown is a magnified version of the object spectrum, by the factor
indicated on the figure, and offset for clarity. The residual spectrum
is shown on top of the object spectrum, and is also magnified by the
factor indicated on the figure. The noise used to compute the
significance of the fits is labeled on the top right. The noise is
the RMS dispersion of the residual spectrum, and is in the same units
as the $y$-axis. The height of the shaded rectangles on top of the
$x$-axis indicates the relative weights used in the baseline
definition.}
  \label{fig:hd23480} 
\end{figure*}

HD\,23480 (= 23\,Tau = Merope) is a well-known member of the Pleiades
cluster and the only target in our list which has a published
$^{12}$C/$^{13}$C-ratio \citep[$41\pm9$,][]{1987ApJ...317..926H}. We
observed it, despite its northern location, as a reference
target. Because of its position the sky, the airmass during our
observations was relatively high. Our result is in disagreement with
\citet[$41\pm9$,][]{1987ApJ...317..926H}: We do not confirm their
low value of $R$ toward HD\,23480.
The {\sc Uves} CH$^+$ spectrum (see Fig.~\ref{fig:hd23480}) appears to be
free of blends or artifacts.

\subsubsection{\object{HD\,35149}}

\begin{figure*}
  \centering
  \includegraphics{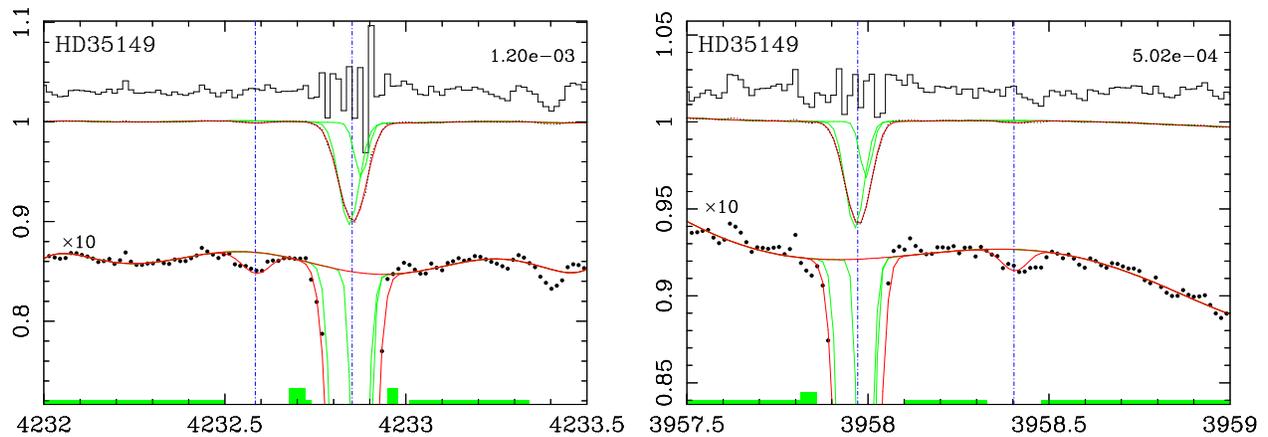}
  \caption{Same as Fig.~\ref{fig:hd23480}, but for HD\,35149}
  \label{fig:hd35149}
\end{figure*}

A detailed study of the interstellar medium towards HD\,35149 (= 23
Ori) based on high resolution UV spectra has been published by
\citet{1999ApJS..124..465W}.  \citet{2003ApJ...595..235A} searched the
star for rotationally resolved C$_3$, but did not detect this molecule
towards HD\,35149. Measurements of diffuse interstellar bands towards the star 
have been published by \citet{2003ApJ...584..339T}.
The {\sc Uves} CH$^+$ spectrum (see Fig.~\ref{fig:hd35149}) appears to be
free of blends or artifacts.

\subsubsection{\object{HD\,37903}}

\begin{figure*}
  \centering
  \includegraphics{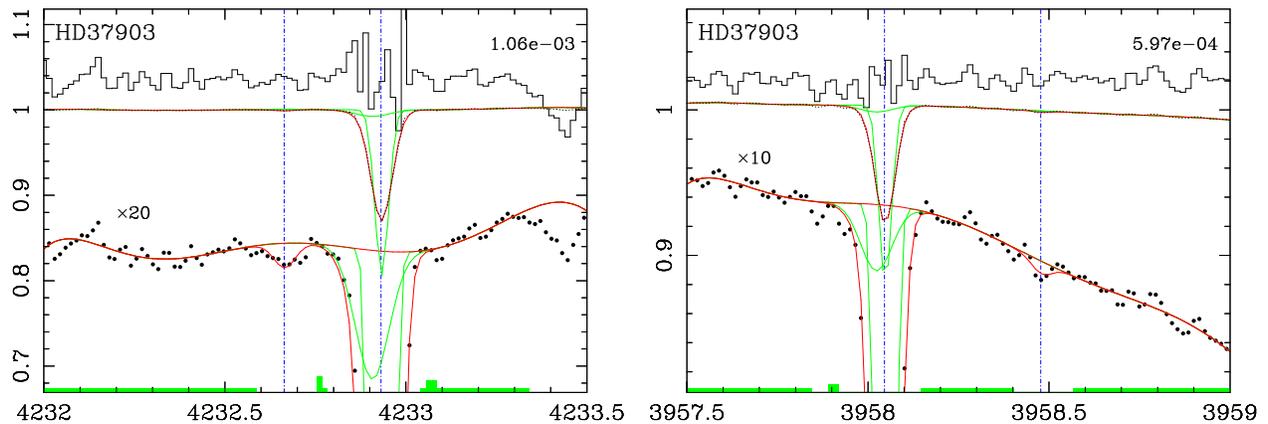}
  \caption{Same as Fig.~\ref{fig:hd23480}, but for HD\,37903}
  \label{fig:hd37903}
\end{figure*}

HD\,37903 is the illuminating star of the reflection nebula NGC\,2023.
A detailed study of the interstellar medium towards HD\,37903, based
on ORFEUS UV data, has been published by \citet{2002ApJ...575..234L}.
This star has previously been observed in the CH$^+$-line by
\citet{1993A&A...269..477G}.
The {\sc Uves} CH$^+$ spectrum (see Fig.~\ref{fig:hd37903}) appears to be
free of blends or artifacts.

\subsubsection{\object{HD\,52266}}

HD\,52266 is considered a field star, but is projected close to CMa
OB1. \citet{2004A&A...425..937D} found evidence for a cluster near the
star. No detailed study of the interstellar medium in this direction
has been published so far.

\begin{figure*}
  \centering
  \includegraphics{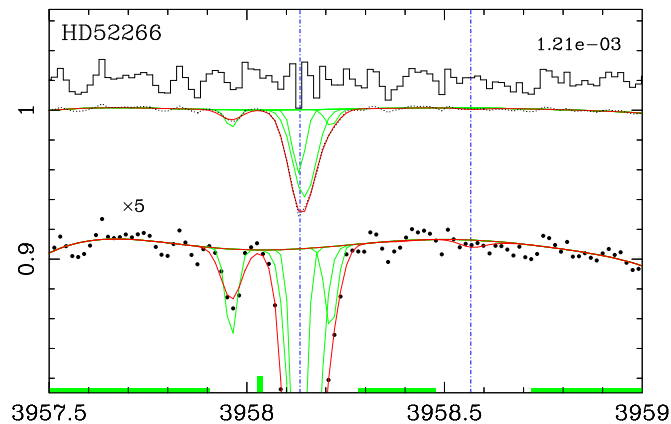}
  \caption{Same as Fig.~\ref{fig:hd23480}, but for HD\,52266}
  \label{fig:hd52266}
\end{figure*}

This spectrum is very shallow because we interrupted the observation
when finding out at the telescope that the line of sight to HD\,52266
is very structured, and useless for the determination of an isotope
ratio. The multiple velocity components are nonetheless faint enough
to have escaped detection in our {\sc Feros} preselection. The {\sc Uves}
CH$^+$ spectrum is shown in Fig.~\ref{fig:hd52266}.

\subsubsection{\object{HD\,52382}}

\begin{figure*}
  \centering
  \includegraphics{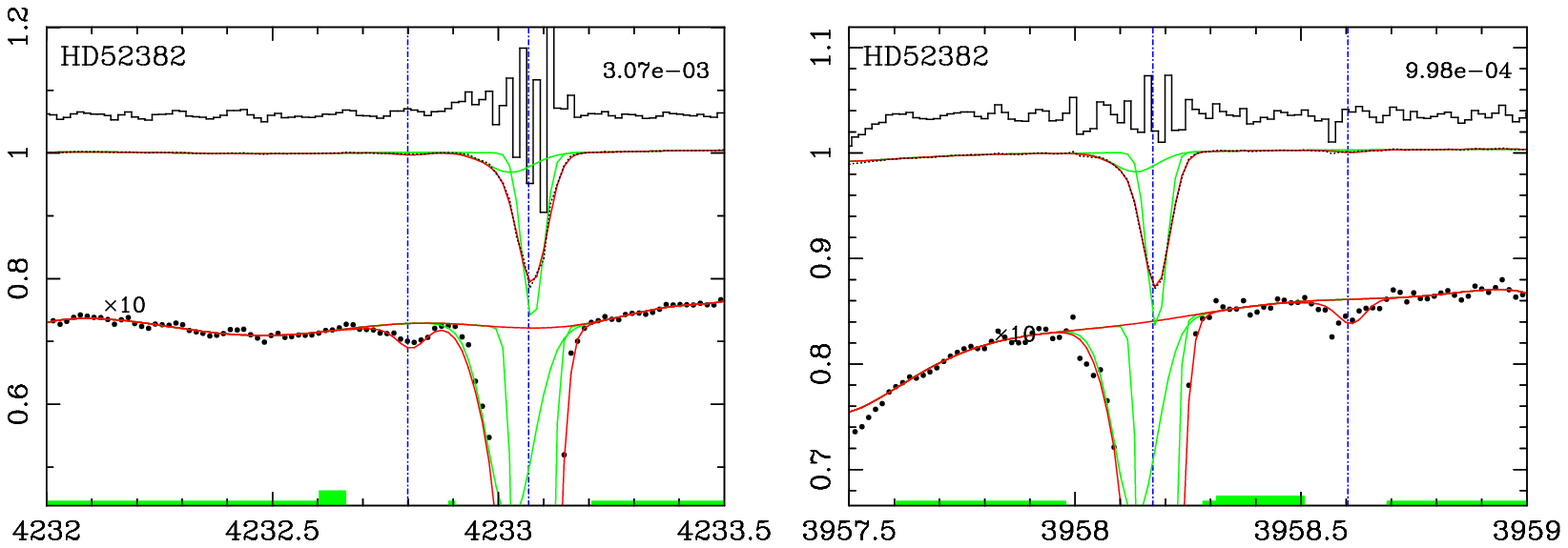}
  \caption{Same as Fig.~\ref{fig:hd23480}, but for HD\,52382}
  \label{fig:hd52382}
\end{figure*}

This star is a member of the CMa OB1 association and has previously
been observed in the CH$^+$-line by \citet{1997A&A...320..929G}.
The {\sc Uves} CH$^+$ spectrum (see Fig.~\ref{fig:hd52382}) appears to be
free of blends or artifacts.

\subsubsection{\object{HD\,53974}}

\begin{figure*}
  \centering
  \includegraphics{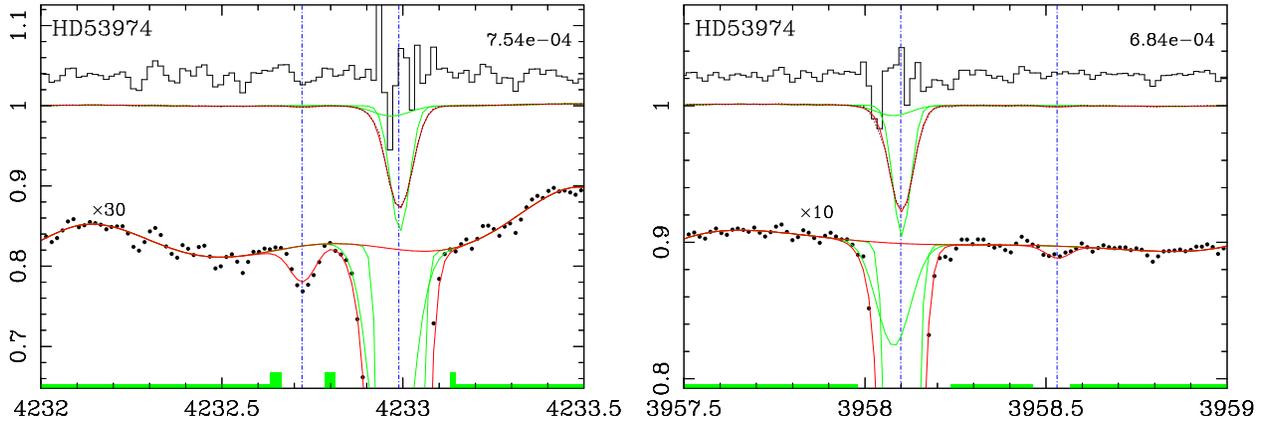}
  \caption{Same as Fig.~\ref{fig:hd23480}, but for HD\,53974}
  \label{fig:hd53974}
\end{figure*}

This star is considered a field Be star by
\citet{2005A&A...441..235Z}. No detailed study of the interstellar
medium in this direction has been published so far.
The {\sc Uves} CH$^+$  spectrum (see
Fig.~\ref{fig:hd53974}) appears to be free of blends or artifacts.

\subsubsection{\object{HD\,58510}}

HD\,58510 is a little studied B1ab supergiant. No detailed study of
the interstellar medium in this direction has been published so far.

\begin{figure*}
  \centering
  \includegraphics{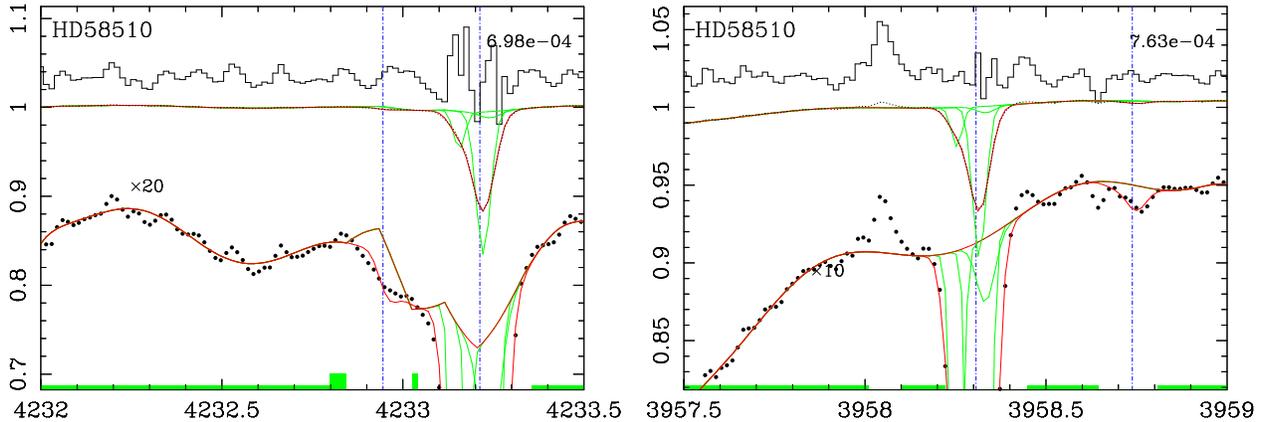}
  \caption{Same as Fig.~\ref{fig:hd23480}, but for HD\,58510}
  \label{fig:hd58510}
\end{figure*}

Although the fundamental $^{13}$CH$^+$ tone is only marginally
blended, the baseline at $\lambda$4232 shows a break right under
$^{13}$CH$^+$, which is very difficult to account for. The {\sc Uves}
CH$^+$ spectrum is shown in Fig.~\ref{fig:hd58510}.

\subsubsection{\object{HD\,73882}}

\begin{figure*}
  \centering
  \includegraphics{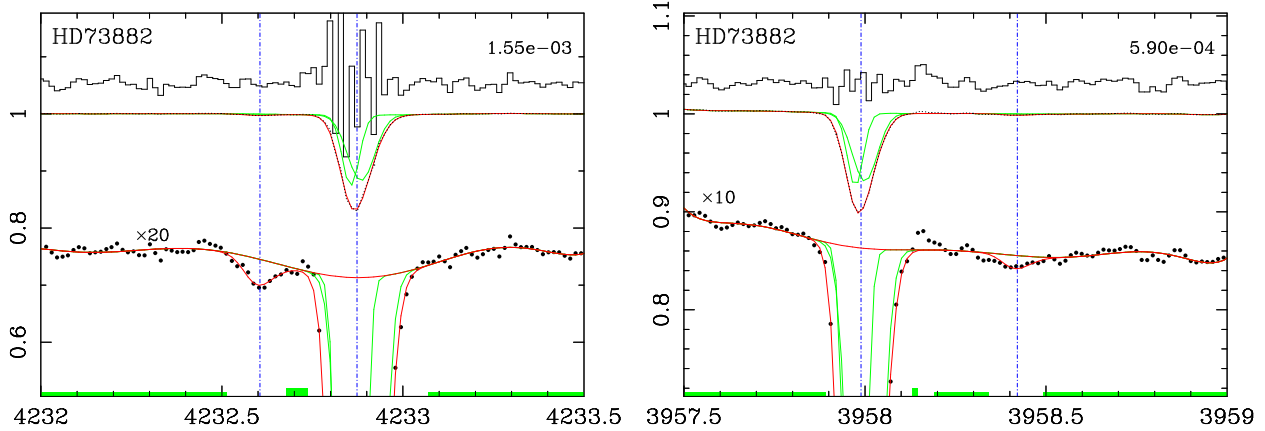}
  \caption{Same as Fig.~\ref{fig:hd23480}, but for HD\,73882}
  \label{fig:hd73882}
\end{figure*}

This member of Vela OB1 has previously been observed in the
CH$^+$-line by \citet{1993A&A...269..477G}. The star is located in the
Gum nebula.  HD\,73882 has been observed in the far ultraviolet
  with FUSE.  These observations have been used to study e.g. the
  interstellar deuterium abundance \citep{2000ApJ...538L..69F} and the
  interstellar extinction towards HD\,73882
  \citep{2005ApJ...625..167S}. FUSE spectra of the star have also been
  published by \citet{2002ApJ...573..662S} and
  \citet{2002ApJ...577..221R}. IUE and FUSE data have been used by
  \citet{2007ApJS..168...58S} to determine a $^{12}$CO/$^{13}$CO
  column density ratio of 25 towards HD\,73882, significantly below
  our value for $^{12}$C/$^{13}$C. Since \citet{2007ApJS..168...58S} find
  particularly low $^{12}$CO/$^{13}$CO-ratios in lines of sight which
  sample cool and dense gas, they interpret this finding as evidence
  for chemical fractionation of CO in cool and dense regions. The
{\sc Uves} CH$^+$ spectrum (see Fig.~\ref{fig:hd73882}) appears to be
free of blends or artifacts.

\subsubsection{\object{HD\,75149}}

\begin{figure*}
  \centering
  \includegraphics{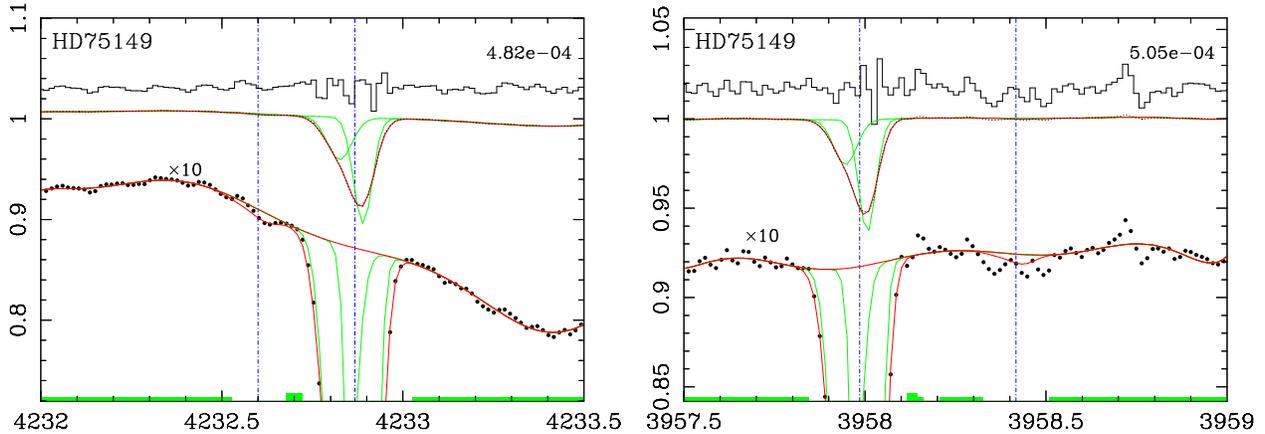}
  \caption{Same as Fig.~\ref{fig:hd23480}, but for HD\,75149}
  \label{fig:hd75149}
\end{figure*}

This star is a member of the Vela OB1 association and has previously
been observed in the CH$^+$-line by
\citet{1997A&A...320..929G}. \citet{2002A&A...389..993G} and
\citet{1999A&A...351..657G} also observed the star in the interstellar
CN and C$_2$ lines, respectively. Observations of interstellar Na\,{\sc
i} and Ca\,{\sc ii} have been published by \citet{2000ApJS..126..399C}.
The {\sc Uves} CH$^+$ spectrum (see Fig.~\ref{fig:hd75149}) appears to be
free of blends or artifacts.

\subsubsection{\object{HD\,76341}}

Although this star is not in the list of \citet{1978ApJS...38..309H},
it is considered by \citet{2000AJ....119.1855R} as a probable member
of the OB association Vela OB1.  \citet{2004ApJ...610..285N} searched
for unusually strong excited interstellar C\,{\sc i} in stars seen
through the Vela supernova remnant, indicating shock excitation, and
found several stars with unusually strong C\,{\sc i}. Both HD\,35149
and HD\,76341 are included in their sample, but do not show unusually
strong C\,{\sc i}.

Although relatively faint, HD\,76341 gives us the best measurement of
$^{12}$CH$^+$/$^{13}$CH$^+$. It has a very sharp and relatively strong
CH$^+$-line.

\begin{figure*}
  \centering
  \includegraphics{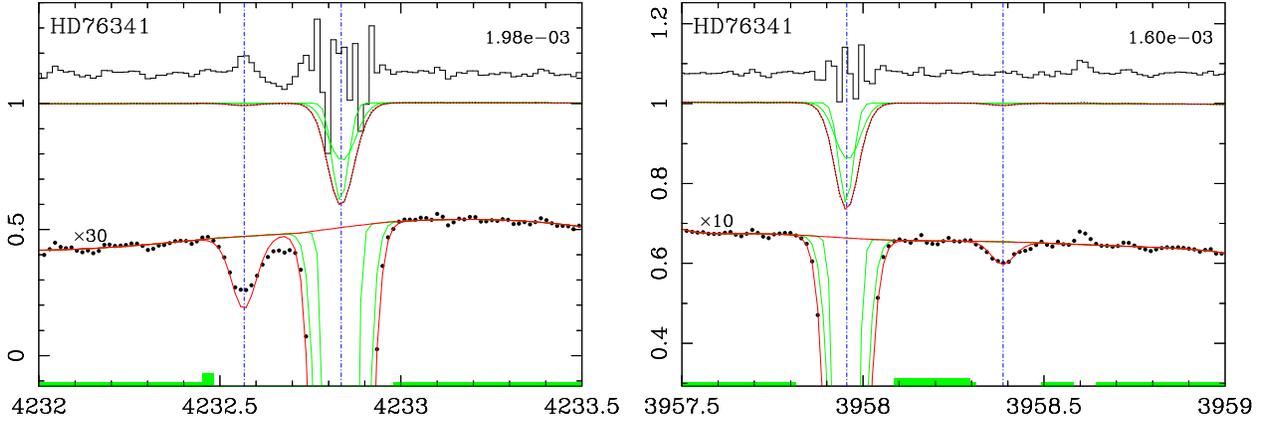}
  \caption{Same as Fig.~\ref{fig:hd23480}, but for HD\,76341}
  \label{fig:hd76341}
\end{figure*}

The fundamental $^{13}$CH$^+$ absorption at $\lambda$4232 appears to
be blended with the main isotope, as well as affected by baseline
uncertainties (atmospheric, and instrumental detector glitches). By
artificially changing the noise under $\lambda$3957 and $\lambda4232$,
we can test the effect of imposing the $R$ value from each
overtone. We find it is impossible to fit the $^{13}$CH$^+\lambda$4232
simultaneously with $^{12}$CH$^{+}\lambda$4232 without changing the
baseline under $^{13}$CH$^{+}\lambda$4232 by deviations from the
Legendre polynomial fit (step~\ref{it:local}) comparable to the
$^{13}$CH$^{+}$ line depth.  On the other hand, if we artificially
decrease the noise under the overtone $^{13}$CH$^+$ absorption at
$\lambda$3957, then it is possible to also fit $^{12}$CH$^{+}$ in both
overtones. Fig.~\ref{fig:hd76341} shows the result of imposing the
$^{13}$CH$^+\lambda$3957 profile. It is clear that the observed
spectral points in $^{13}$CH$^{+}\lambda$4232 do not share the same
profile as the other 3 lines, $^{13}$CH$^{+}\lambda$4232 is therefore
affected by either atmospheric features or detector glitches, of the
type seen at $\lambda = 3858.6$\,\AA\@. The uncertainties on $R$ are
those obtained by varying only $R$ and keeping fixed all other
parameters. The {\sc Uves} CH$^+$ spectrum is shown in Fig.~\ref{fig:hd76341}.

\subsubsection{\object{HD\,91452}}

\begin{figure*}
  \centering
  \includegraphics{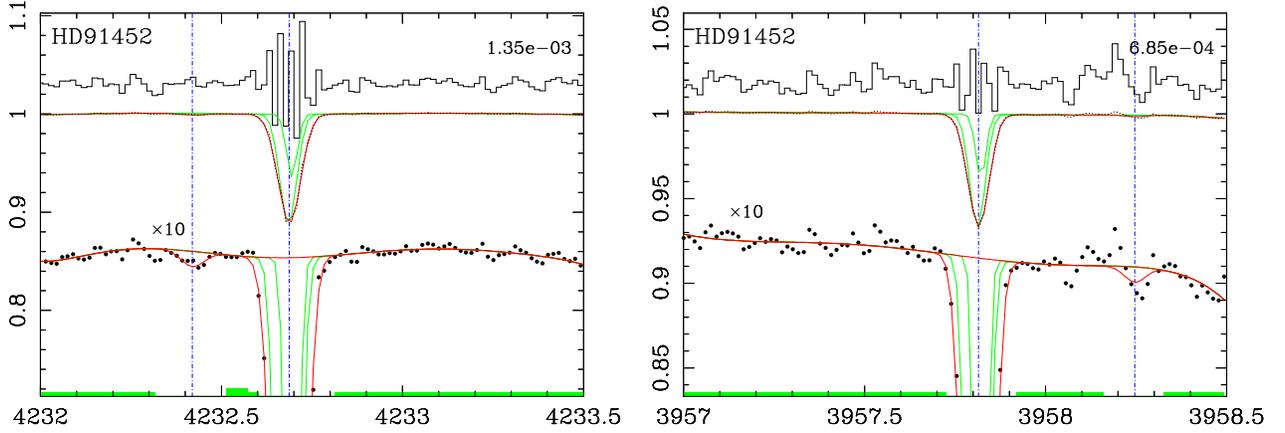}
  \caption{Same as Fig.~\ref{fig:hd23480}, but for HD\,91452}
  \label{fig:hd91452}
\end{figure*}

\citet{2004A&A...425..937D} unsuccessfully searched for a cluster
near HD\,91452.  According to \citet{2005A&A...437..247D}, HD\,91452
is a runaway O-type star. No detailed study of the interstellar
medium in this direction has been published so far.
The {\sc Uves} CH$^+$ spectrum (see Fig.~\ref{fig:hd91452}) appears to be
free of blends or artifacts.

\subsubsection{\object{HD\,92964}}

\begin{figure*}
  \centering
  \includegraphics{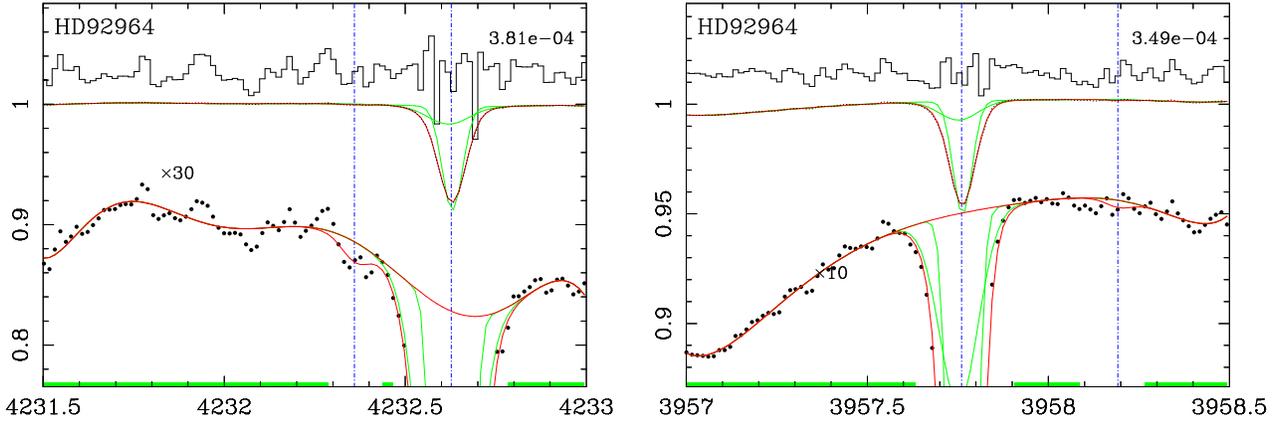}
  \caption{Same as Fig.~\ref{fig:hd23480}, but for HD\,92964}
  \label{fig:hd92964}
\end{figure*}

Member of the OB association Car OB1. No detailed study of the
interstellar medium in this direction has been published so far.
The {\sc Uves} CH$^+$ spectrum (see Fig.~\ref{fig:hd92964}) appears to be
free of detectable blends or artifacts.

\subsubsection{$\zeta$\,Oph}

\begin{figure*}
\begin{center}
\includegraphics{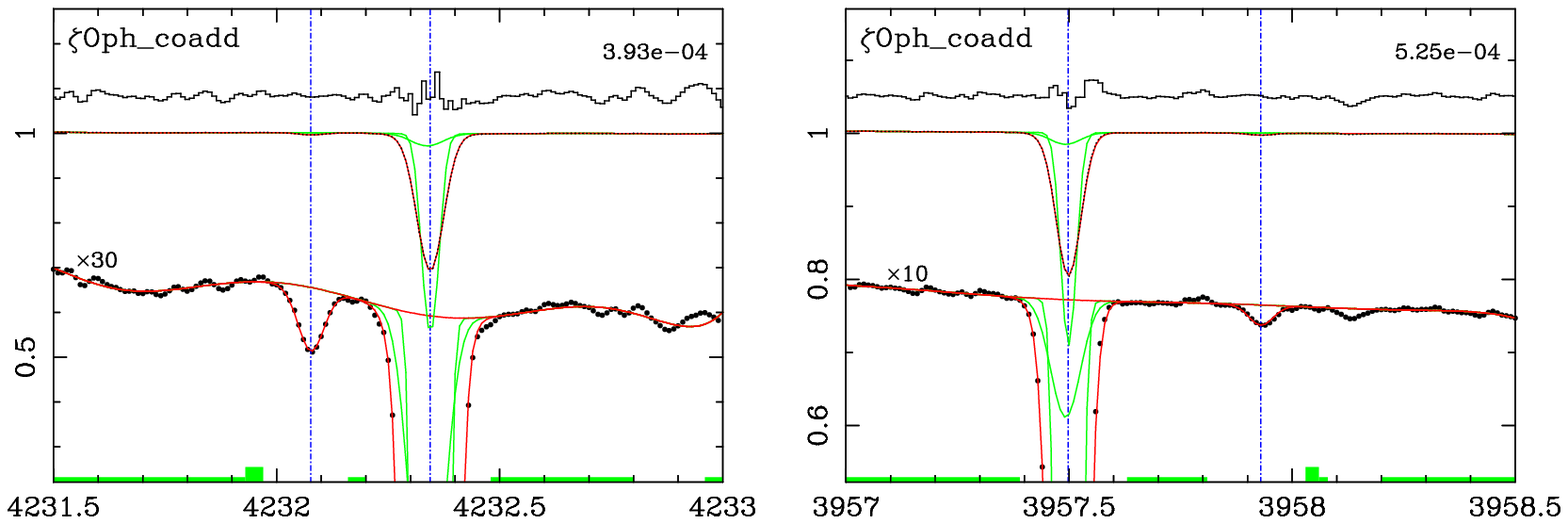}
\end{center}
\caption{Same as Fig.~\ref{fig:hd23480}, but for $\zeta$\,Oph.}
  \label{fig:zetaoph}
\end{figure*}

Because of its brightness and clean interstellar line profiles,
$\zeta$\,Oph is a prototype star for studies of the interstellar medium,
also for the C$^{12}$/C$^{13}$ ratio. The {\sc Uves} CH$^+$ spectrum 
is shown in Fig.~\ref{fig:zetaoph}. For detailed discussion see paper\,I.

\subsubsection{HD~110432}

\begin{figure*}
\begin{center}
\includegraphics{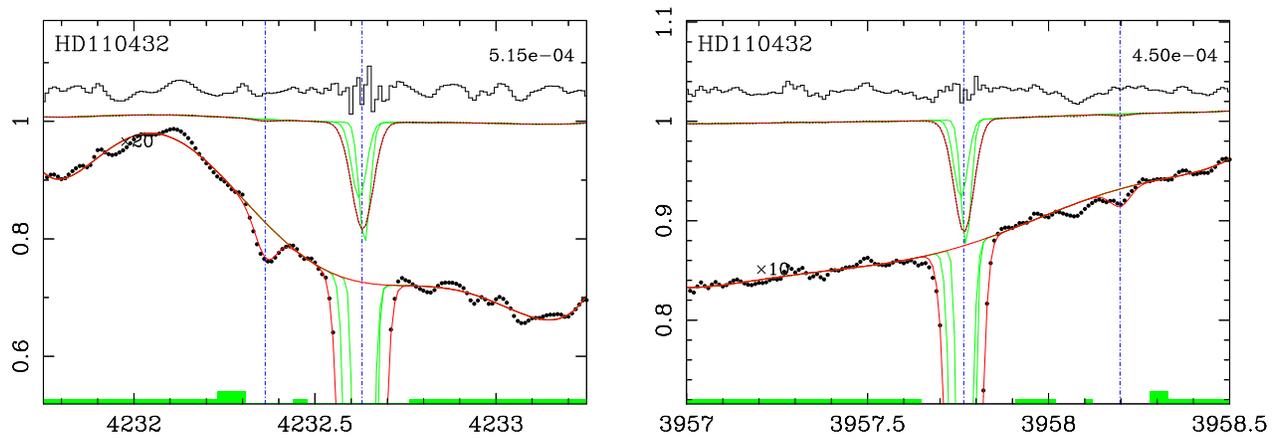}
\end{center}
\caption{Same as Fig.~\ref{fig:hd23480}, but for HD\,110432}
  \label{fig:hd110432}
\end{figure*}

This star is also discussed in paper\,I.  The sight-line to HD\,110432
is one of the cleanest sight-lines studied in paper\,I. The {\sc Uves} CH$^+$
spectrum is shown in Fig.~\ref{fig:hd110432}.

\subsubsection{HD~152235}

\begin{figure*}
\begin{center}
\includegraphics{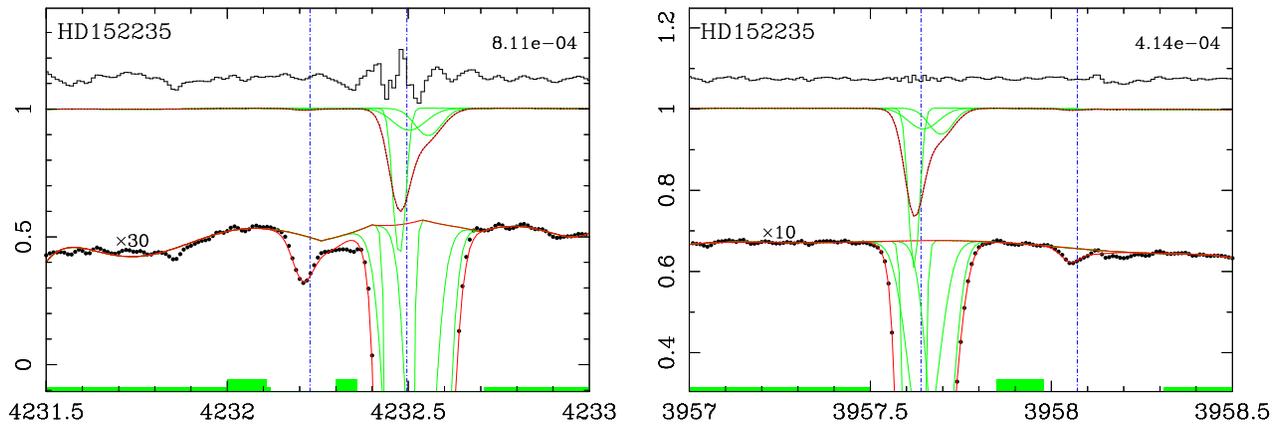}
\end{center}
\caption{Same as Fig.~\ref{fig:hd23480}, but for HD\,152235.}
  \label{fig:hd152235}
\end{figure*}

This star is another star from paper\,I.  We treated this case as a
blend, since the $^{13}$CH$^+$ profile just overlaps with
$^{12}$CH$^+$ in the fundamental tone. We exaggerated the noise for
$\lambda$4232 by a factor of 3, instead of 10 as in the other blended
cases. There is possible telluric absorption red-wards of
$^{13}$CH$^+\lambda$3957. The {\sc Uves} CH$^+$ spectrum is shown Fig.~\ref{fig:hd152235}.

\section{Discussion and conclusions} \label{sec:conc}

Our systematic search for stars with clean line profiles in CH$^+$ and
large equivalent widths, which are suitable for isotope studies, has
resulted in a much increased sample of reliable $^{12}$C/$^{13}$C
isotope ratios from optical observations.

The VLT-{\sc Uves} spectra of CH$^+$ absorption in the lines of sight
towards 14 stars have allowed us to measure the carbon isotopic ratio
$R= ^{12}$C/$^{13}$C with unprecedented accuracy. The significant
extension of the sample of \citet{2005A&A...441..181C} allows us to
strengthen their conclusion that there is truly a significant scatter
in the local ISM\@.  Averaging our measurements for the 14 lines of
sight gives a value representative of the local ISM: $\langle R
\rangle = 76.27 \pm 1.94$.  This mean ratio is in excellent agreement
with the local ratio determined from radio-astronomical measurements
of CO and H$_2$CO \citep[76$\pm$7,][]{1994ARA&A..32..191W}, and also
in good agreement with more recent determinations from CN
\citep[$\approx$68,][]{2005ApJ...634.1126M}. This agreement of results
obtained from different molecules suggests that chemical fractionation
and selective dissociation do not play a major role for the CO,
H$_2$CO and CN molecules. The results of
  \citet{2007ApJS..168...58S} suggest, however, that CO may be
   affected by chemical fractionation  in cool and
  dense regions.

The weighted scatter of $f$ values is $6.3\times\sigma(\langle f
\rangle)$ which directly shows that the $^{12}$C/$^{13}$C isotope
ratio is variable on the relatively small spatial scales which we
sample.  Considering this intrinsic scatter of the $^{12}$C/$^{13}$C
ratio in the ISM, the solar carbon isotope ratio of $89$ is
undistinguishable from the present-day value, even after 4.5\,Gyr of
galactic evolution. This statement is not affected by instrumental
uncertainties, although a sampling bias could distort our ISM value.
Of course, this conclusion is based on the assumption that the solar
system ratio is representative for the galacto-centric distance
R$_\odot$ = 8.5 kpc, 4.5\,Gyr ago. 
Since we have shown that the interstellar $^{12}$C/$^{13}$C isotope
ratio is spatially variable now, it is not clear if the solar value is
characteristic for the local $^{12}$C/$^{13}$C ratio, 4.5\,Gyr
ago. However, there is also no strong evidence that the solar value is
untypical: Although it has been claimed that the sun is overabundant,
especially in carbon and oxygen, compared to the local ISM, the solar
carbon and oxygen abundances have been revised downwards.  Recent
papers \citep[e.g.][]{arXiv:astro-ph/0702429v1} find good agreement
for the carbon and oxygen abundance in the sun, the local ISM and
young stars. 
Comets show a $^{12}$C/$^{13}$C ratio consistent with the photospheric
solar value \citep[e.g.][]{2005A&A...432L...5M}. Therefore it seems
also unlikely that the $^{12}$C/$^{13}$C ratio was modified from the
value in the solar nebula by fractionation of carbon isotopes during
the formation of the sun. 

Therefore we conclude that there is no evidence that the solar
$^{12}$C/$^{13}$C ratio is untypical for the solar neighborhood 4.5
Gyr ago. It is plausible however, that the average value for the solar
neighborhood was somewhat different from the solar value by a similar
amount as the present-day variations. A small decrease of the average
$^{12}$C/$^{13}$C ratio in the solar neighborhood during the lifetime
of the sun is therefore certainly possible.

The $^{12}$C/$^{13}$C ratio increases with galacto-centric distance
\citep{1994ARA&A..32..191W} by about 7.5$\pm$1.9 per kpc. Therefore, a
comparison of the present-day local $^{12}$C/$^{13}$C ratio with the
solar photospheric value should also take into account that the sun
was born at a different galacto-centric distance
\citep{1997A&A...326..139W}. Since the sun has, according to
\citet{1996A&A...314..438W}, migrated from its birth-place in the
inner part of the Galaxy outwards by $1.9\pm0.9$ kpc during its
lifetime, this effect decreases the expected change in the
$^{12}$C/$^{13}$C ratio significantly.

Early models for the $^{12}$C/$^{13}$C isotope ratio were able to
model the significant decrease of the local ratio over the lifetime of
the sun (about a factor two), which was thought at that time to be
present. However, the models require some fine tuning and calibration,
so that a smaller decrease is not in obvious conflict with the models.
Nevertheless, some variation of the $^{12}$C/$^{13}$C isotope ratio is
to be expected. New models of the chemical evolution of the Galaxy are
required to take into account the new observational results.

A larger sample of stars will allow a better estimate of the average
ISM value. In addition, these observations will allow to determine the
scale of the variations of the isotope ratio and thus the efficiency
of mixing on different scales.

\begin{acknowledgements}
  S.C. acknowledges support from Fondecyt grant 1030805, and from the
  Chilean Center for Astrophysics FONDAP 15010003.  This research has
  made use of the SIMBAD database, operated at CDS, Strasbourg, France.
\end{acknowledgements}

\bibliographystyle{aa}
\bibliography{c12c13}

\end{document}